\documentclass[12pt]{article}

\usepackage{color}
\usepackage{epsfig,amssymb,amsfonts,amsmath,graphicx,dsfont,cite,xfrac}
\usepackage{array}
\usepackage{subfigure}
\usepackage{setspace}
\usepackage{rotating}
\usepackage{pdflscape}

\usepackage{authblk}
\definecolor{mygray}{gray}{0.5}

\usepackage{cite}
\usepackage[colorlinks=true,linkcolor=blue,citecolor=red]{hyperref}

\newcommand{\be}{\begin{equation}}
\newcommand{\ee}{\end{equation}}
\newcommand{\bea}{\begin{eqnarray}}
\newcommand{\eea}{\end{eqnarray}}

\title{An Integro-Differential Equation of the Fractional Form: Cauchy Problem and Solution}

\author[${}$]{Fernando Olivar-Romero}
\author[${}$]{Oscar Rosas-Ortiz}

\affil[${}$]{\footnotesize Physics Department, Cinvestav, AP 14-740, 07000
M\'exico City, Mexico}

\date{}
\begin{document}

\maketitle

\begin{abstract}
We solve the Cauchy problem defined by the fractional partial differential equation $[\partial_{tt}-\kappa\mathbb{D}]u=0$, with $\mathbb{D}$ the pseudo-differential Riesz operator of first order, and the initial conditions $u(x,0)=\mu(\sqrt{\pi}x_0)^{-1}e^{-(x/x_0)^2}$, $u_t(x,0)=0$. The solution of the Cauchy problem resulting from the substitution of the Gaussian pulse $u(x,0)$ by the Dirac delta distribution $\varphi(x)=\mu\delta(x)$ is obtained as corollary.
\end{abstract}


\section{Introduction}

Linear partial differential equations of second order are useful in physics to model phenomena like wave propagation, heat diffusion and transport processes \cite{Tik63,Duf15,Bor18}. In analogy to conics of analytic geometry,  the wave equation is hyperbolic while the heat and transport equations are parabolic. In a recent work \cite{Oli18} we have reported a fractional formulation that permits the study of such equations in unified form. Additionally, we have introduced an integro-differential version of the parabolic equation $u_{tt}-\kappa u_{x}=0$ (hereafter called {\em complementary equation}) that is solvable in analytic form. That is, in \cite{Oli18} we have solved the Cauchy problem for $u_{tt} -\kappa \mathbb D u =0$ with zero initial velocity and the Dirac delta pulse $\varphi(x)=\mu\delta(x)$ as initial condition. The symbol $\mathbb D$ stands for the pseudo-differential Riesz operator \cite{Rie49} (for contemporary notions on the matter see e.g. \cite{Uma15}). In the present work we provide the solutions for the Cauchy problem with zero initial velocity and the Gaussian distribution $u(x,0)=\mu(\sqrt{\pi}x_0)^{-1}e^{-(x/x_0)^2}$ as initial disturbance.
 
The manuscript is structured as follows. In Section \ref{sec:2} we give the solution of the Cauchy problem for the modified complementary equation when the Gaussian distribution is considered as initial condition with zero initial velocity. We recover the results reported in \cite{Oli18} as a byproduct. In Section \ref{sec:3} we analyze the results. Some final conclusions are given in Section \ref{sec:4}.

\section{Statement of the problem and solution}
\label{sec:2}

The main results of this contribution are summarized in the following Proposition and Corollary.

\vskip1ex
\textbf{Proposition~1.} The Cauchy problem defined for the integro-differential equation
\be 
\frac{\partial^2}{\partial t^2}u(x,t)+\frac{\kappa}{\pi}\frac{\partial}{\partial x}\int_{\mathbb{R}}\frac{u(y,t)}{x-y}dy=0, \quad \kappa >0,
\label{ideq}
\ee
with the initial conditions
\be 
u(x,0)=\frac{\mu}{ x_0 \sqrt{\pi} } e^{-(x/x_0)^2}, \quad u_t(x,0)=0, \quad x_0\geq0, 
\label{gauss} 
\ee
is solved by the function
\be
u(x,t)  =   \displaystyle\frac{\mu}{\kappa t^2} \sum_{k=0}^{\infty} \frac{(-1)^k }{k!} \left( \frac{x_0}{ 2 \kappa t^2} \right)^{2k} \theta_k(x,t),
\label{propo}
\ee
where $\theta_k$ is the following Fox $H$-function,
\be
\theta_k(x,t) = H_{3,3}^{2,1} \left[ \left. \frac{ \vert x \vert }{ \kappa t^2 } \right\rvert 
\begin{array}{c} 
\left( -2k,1 \right),
(\frac12, \frac12),
\left( -1-4k,2 \right)\\[1.5ex]
(0,1), \left(-2k,1 \right), 
(\frac12, \frac12) 
\end{array}
\right].
\label{propo2}
\ee

\textbf{Proof.}  First note that Equation (\ref{ideq}) is indeed the fractional partial differential equation
\be 
\left[\frac{\partial^2}{\partial t^2}-\kappa\mathbb{D}\right]u(x,t)=0, 
\label{ceq} 
\ee
with $\mathbb D$ the pseudo-differential Riesz operator \cite{Rie49,Uma15}. We may consider a generalized version of the latter equation \cite{Oli18,Gor00}, defined as
\be 
[D^\alpha-v_{\alpha,\beta}^2\mathbb{D}^\beta] u(x,t)=0, \hspace{0.5cm}1\leq\alpha\leq2,\hspace{0.5cm}1\leq\beta\leq2, 
\label{general} 
\ee
where the fractional time-derivative $D^\alpha$ is taken in the sense of Caputo \cite{Cap67} (see also \cite{Uma15}), and $\mathbb{D}^\beta$ is the Riesz operator of order $\beta$. In \cite{Oli18} we had already solved Equation (\ref{general}) for the initial conditions (\ref{gauss}). The solution is written as the series
\be
u(x,t)= \frac{\mu}{ \beta t^{\frac{\alpha}{\beta} } \, v_{\alpha,\beta}^{2/\beta} } \sum_{k=0}^{\infty} \frac{(-1)^k }{k!} \left( \frac{x_0}{ 2 t^{\frac{\alpha}{\beta} } \, v_{\alpha,\beta}^{2/\beta} } \right)^{2k}  \Theta_k(x,t;\alpha,\beta),
\label{solg1}
\ee
with
\bea
\Theta_k(x,t;\alpha,\beta) = H_{3,3}^{2,1} \left[ \left.   \frac{ \vert x \vert }{ t^{\frac{\alpha}{\beta} } \, v_{\alpha,\beta}^{2/\beta}  } \right\rvert 
\begin{array}{cc} 
\left( \frac{\beta -(1 + 2k)}{\beta}, \frac{1}{\beta} \right),
(\frac12, \frac12),
\left( \frac{\beta-\alpha (1+2k) }{\beta}, \frac{\alpha}{\beta} \right)\\[1.5ex]
(0,1), \left( \frac{\beta -(1+2k) }{\beta}, \frac{1}{\beta} \right), (\frac12, \frac12) 
\end{array}
\right].
\label{solg2}
\eea
Here  
\begin{multline} 
H_{p,q}^{m,n}\left[x\biggr\rvert \begin{array}{cc} (a_{1},\alpha_{1}),..., (a_{p},\alpha_{p}) \\ (b_{1},\beta_{1}),..., (a_{p},\beta_{p})  \end{array}\right] \\[2.5ex]
=\frac{1}{2\pi i}\int_{L}\frac{\prod_{j=1}^{m}\Gamma(b_{j}+\beta_{j}z)\prod_{i=1}^{n}\Gamma(1-a_{i}-\alpha_{i}z)x^{-z}dz}{\prod_{i=n+1}^{p}\Gamma(a_{i}+\alpha_{i}z)\prod_{j=m+1}^{q}\Gamma(1-b_{j}-\beta_{j}z)}
\nonumber
\end{multline}
is the Fox $H$-function \cite{Kil04,Mat10} for which the labels $m,n,p$, and $q$ are integers such that $0\leq m\leq q$, and $0\leq n\leq p$. Besides $a_{i},b_{j}\in\mathbb{C}$ and $\alpha_{i},\beta_{j}\in(0,\infty)$. 

The solution to the Cauchy problem (\ref{ideq})-(\ref{gauss}) is obtained by evaluating Equations (\ref{solg1}) and (\ref{solg2}) at the point $(\alpha,\beta)=(2,1)$, with $\kappa=v_{2,1}^2$ and $\Theta_k(x,t; 2,1) = \theta_k(x,t)$
$\diamond$.

\vskip1ex
\textbf{Corollary.} If the Gaussian profile of the initial condition $u(x,0)$ of Proposition~1 is substituted by the Dirac delta distribution $\varphi(x) = \mu \delta(x)$, then the solution is given by
\be 
u^{(\delta)} (x,t)=  \left( \frac{\mu}{\kappa t^2} \right)
H_{3,3}^{2,1} \left[ \left. \frac{ \vert x \vert }{ \kappa t^2 } \right\rvert 
\begin{array}{c} 
\left( 0,1 \right),
(\frac12, \frac12),
\left( -1,2 \right)\\[1.5ex]
(0,1), \left(0,1 \right), 
(\frac12, \frac12) 
\end{array}
\right].
\label{udelta}
\ee

\textbf{Proof.} The delta pulse $\varphi(x)$ is recovered from the Gaussian distribution $u(x,0)$ at the limit $x_0 \rightarrow 0$. The proof is simple by noticing that, with the exception of the term with $k=0$, the coefficients of (\ref{propo}) become zero at such a limit. Therefore, $u(x,t) \rightarrow u^{(\delta)} (x,t)$ as $x_0 \rightarrow 0$ $\diamond$.

\section{Analysis of the results} 
\label{sec:3}

\begin{landscape}
\begin{figure}
\centering
\subfigure[$x_0=1, t=0.1$]{
\includegraphics[scale=0.2]{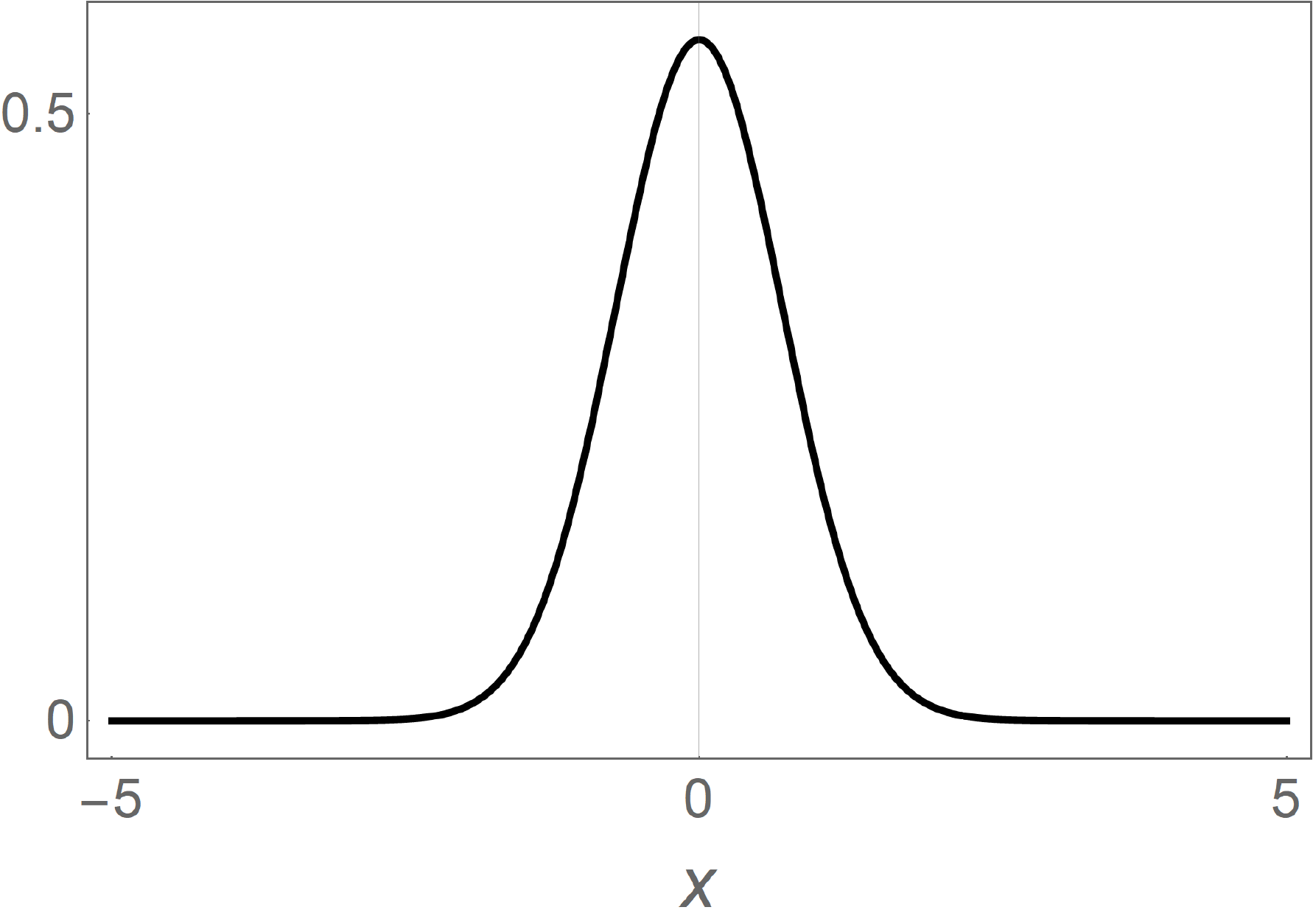} \label{VF-OR-01} }
\hskip2ex
\subfigure[$x_0=1, t=1.7$]{
\includegraphics[scale=0.2]{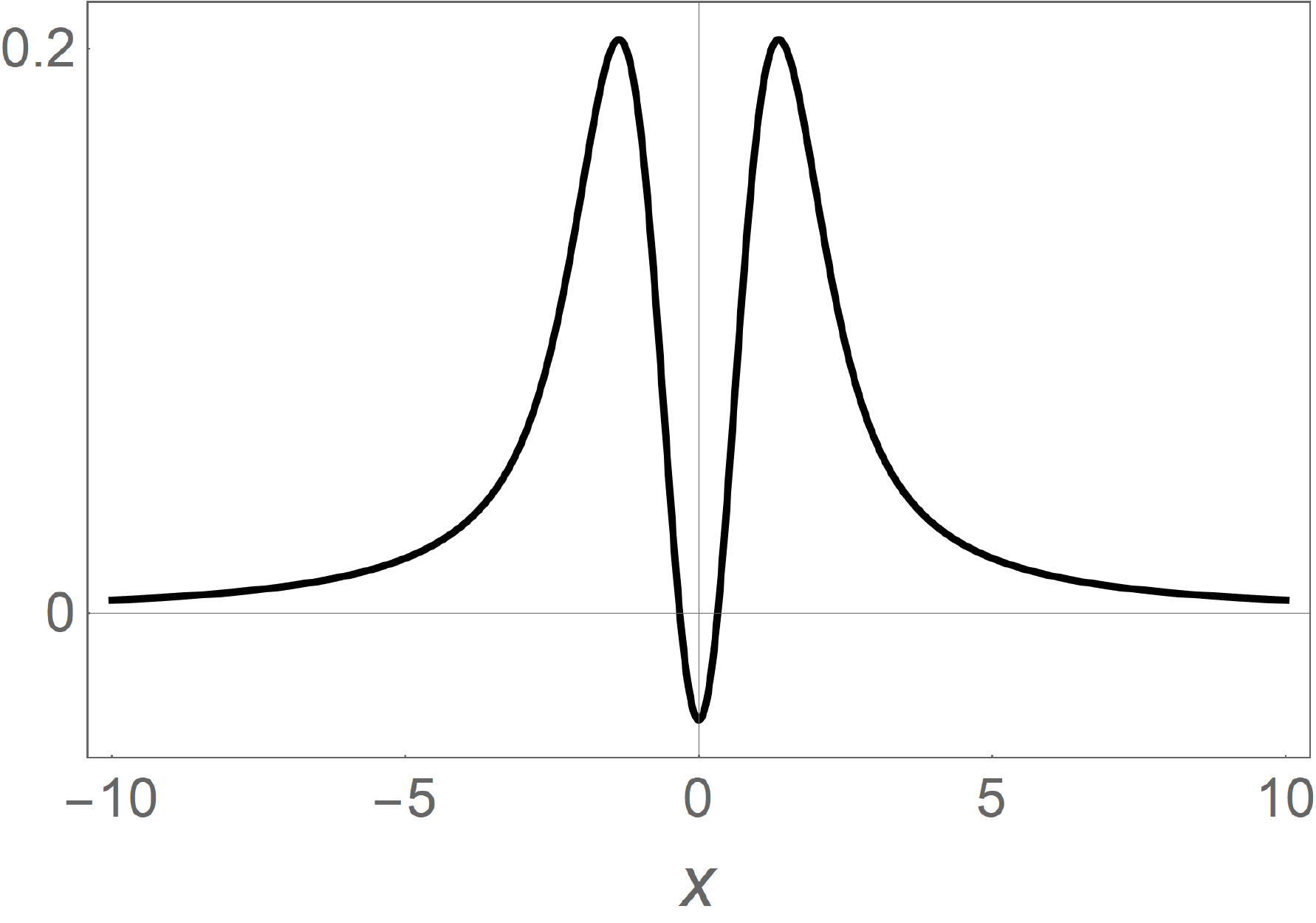} \label{VF-OR-02} }
\hskip2ex
\subfigure[$x_0=1, t=5$]{
\includegraphics[scale=0.2]{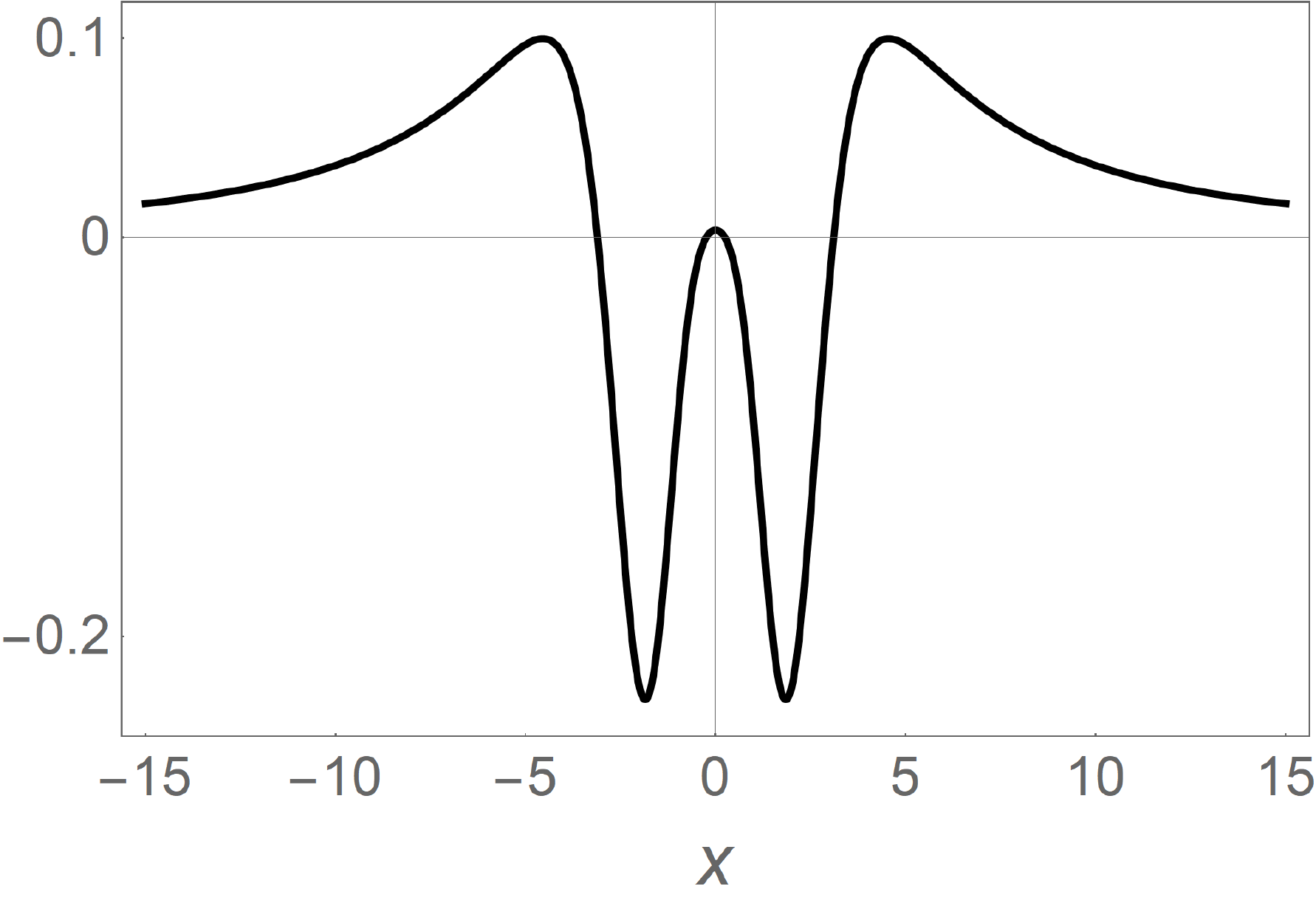} \label{VF-OR-03} }
\hskip2ex
\subfigure[$x_0=1, t=6.5$]{
\includegraphics[scale=0.2]{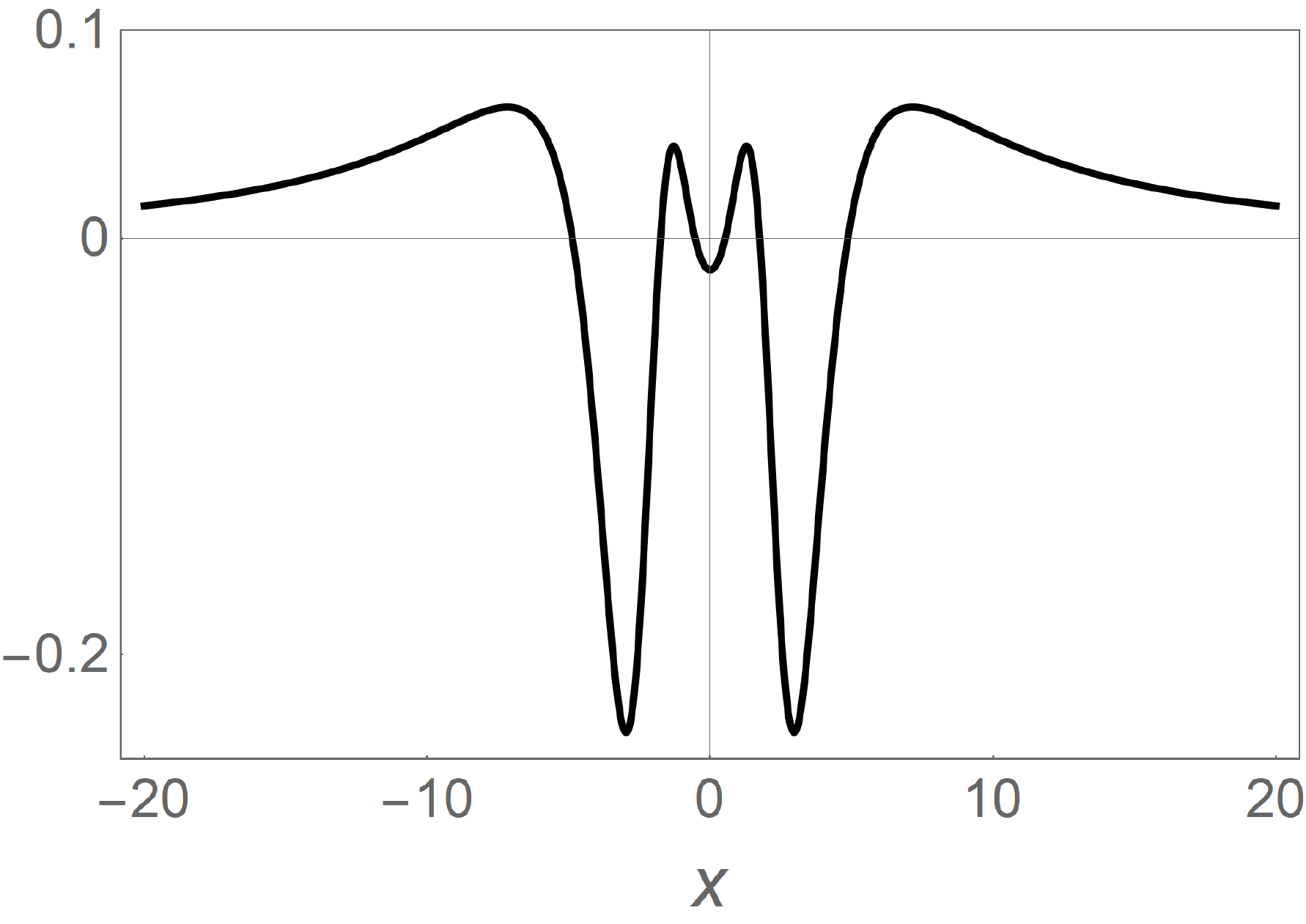} \label{VF-OR-04} }
\vskip2ex 
\subfigure[$x_0=\sqrt{0.5}, t=0.1$]{
\includegraphics[scale=0.2]{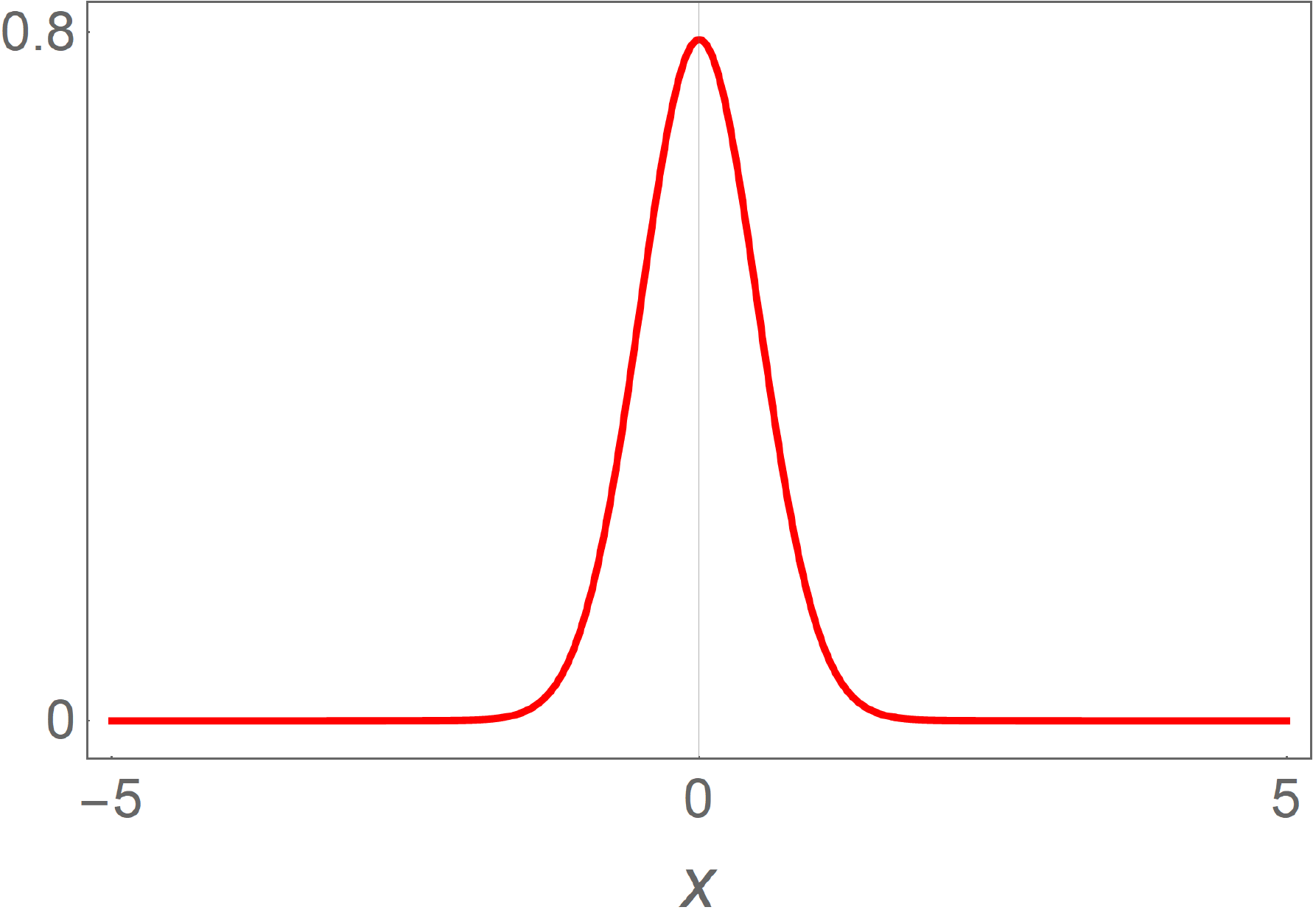} \label{VF-OR-05} }
\hskip2ex
\subfigure[$x_0=\sqrt{0.5}, t=1.7$]{
\includegraphics[scale=0.2]{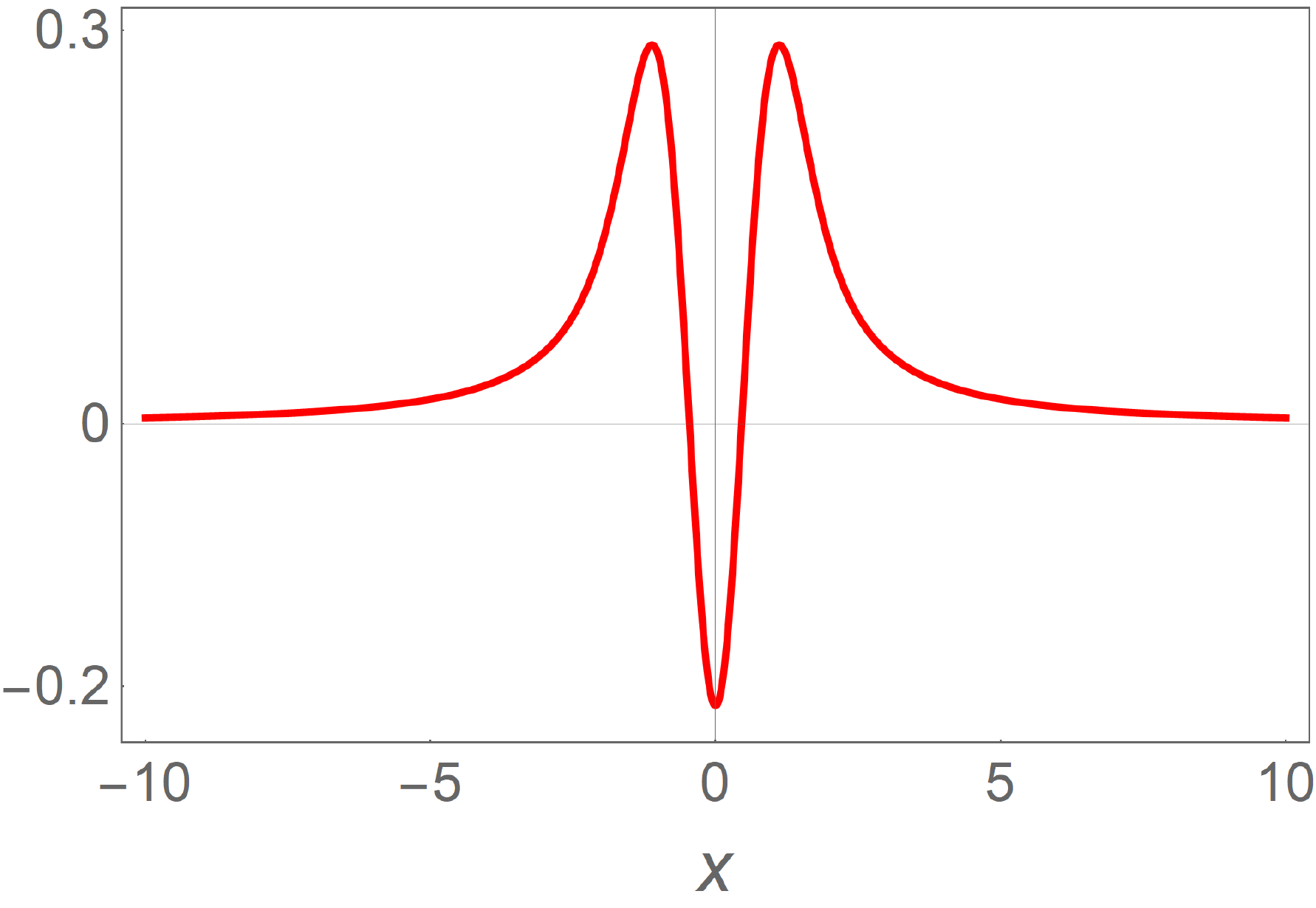} \label{VF-OR-06} }
\hskip2ex
\subfigure[$x_0=\sqrt{0.5}, t=5$]{
\includegraphics[scale=0.2]{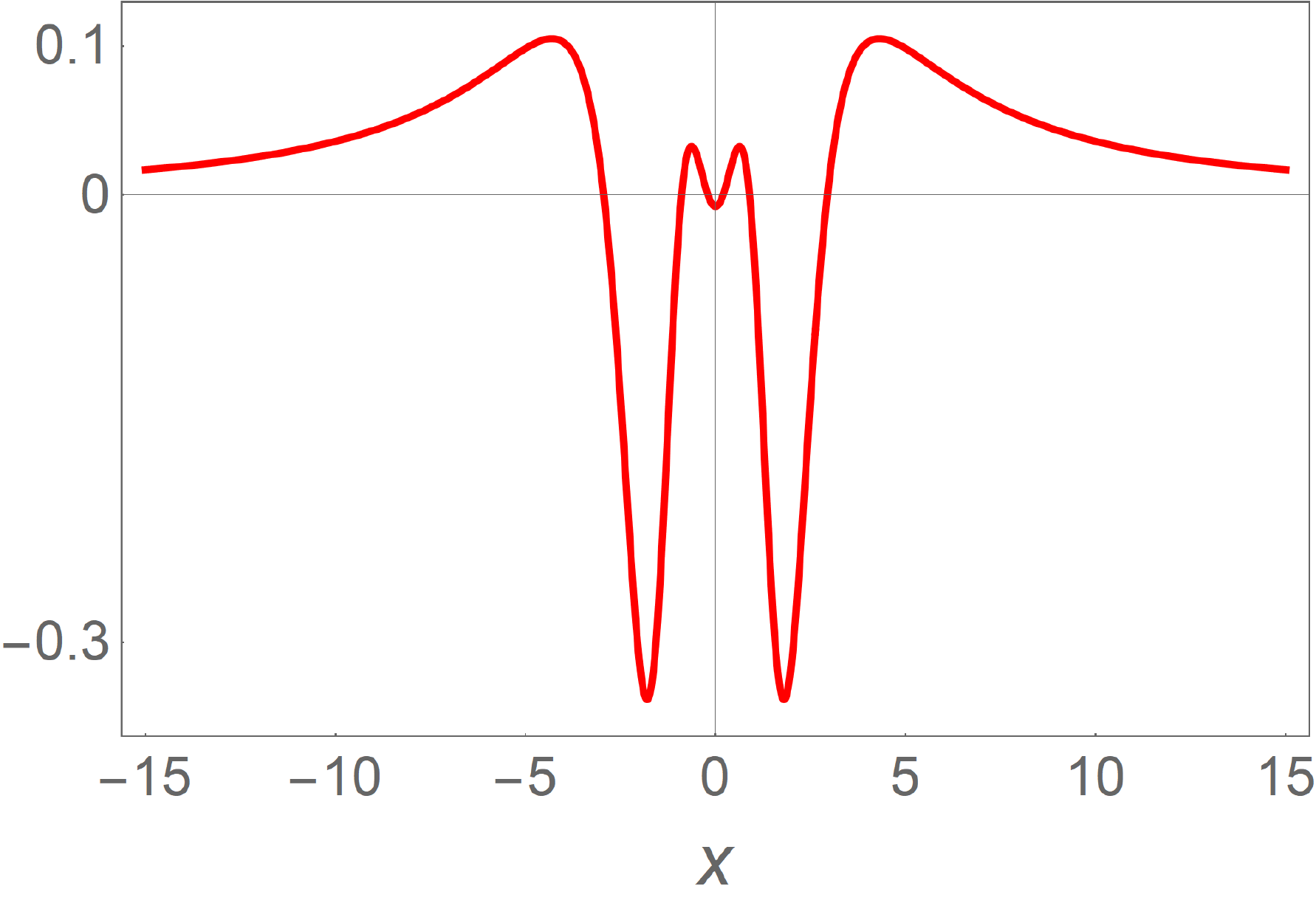} \label{VF-OR-07} }
\hskip2ex
\subfigure[$x_0=\sqrt{0.5}, t=6.5$]{
\includegraphics[scale=0.2]{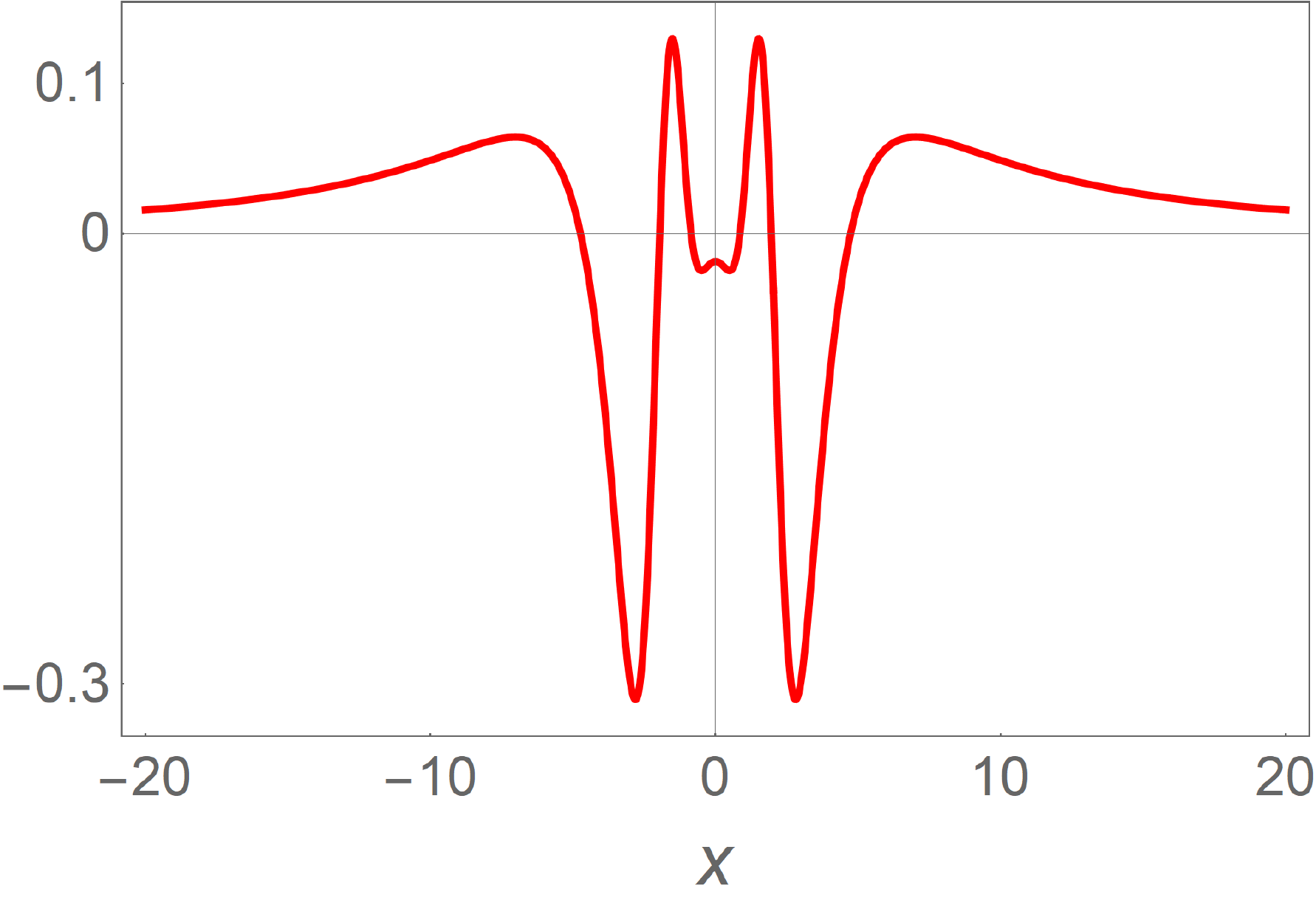} \label{VF-OR-08} }
\vskip2ex 
\subfigure[$x_0=\sqrt{0.1}, t=0.1$]{
\includegraphics[scale=0.2]{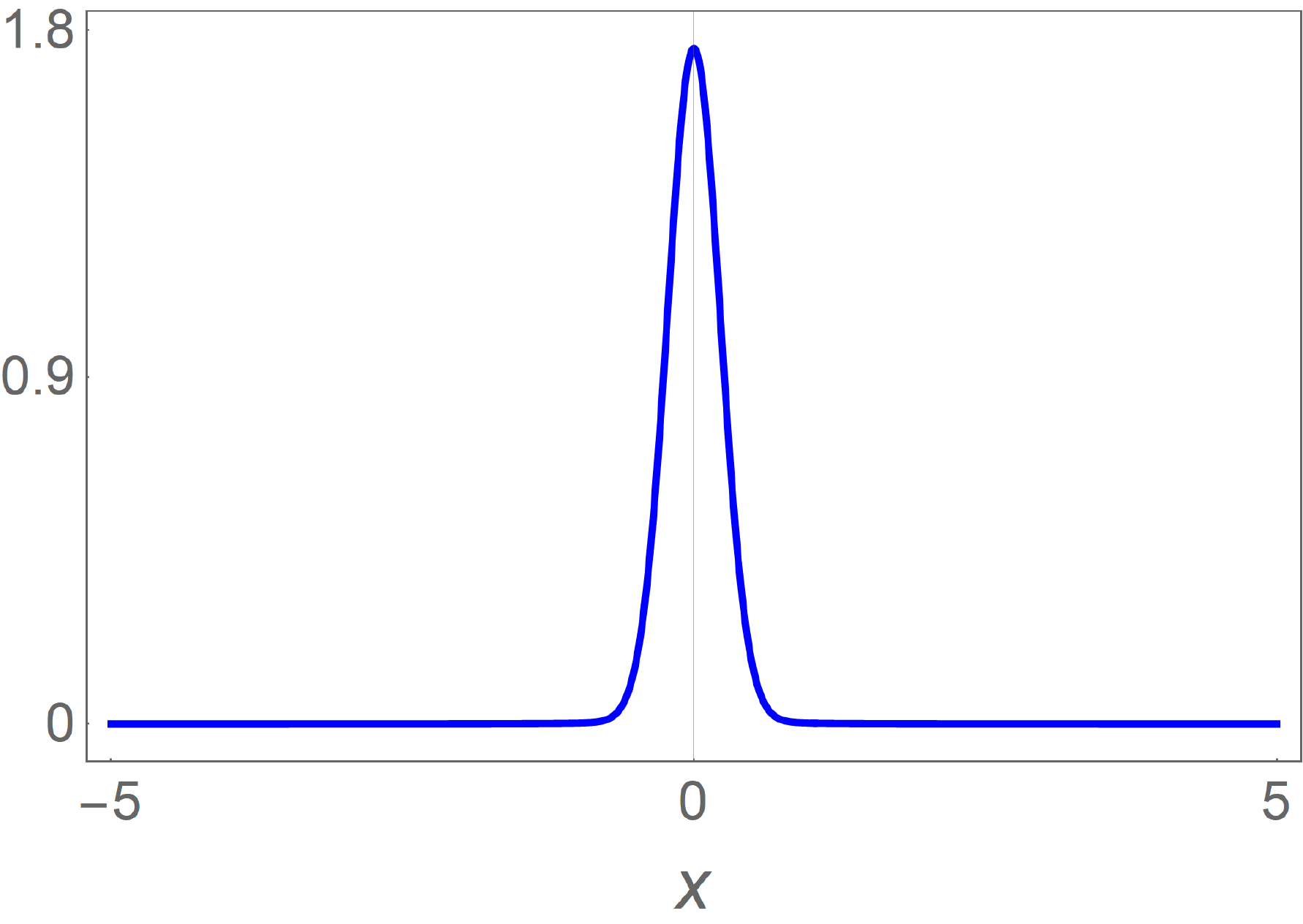} \label{VF-OR-09} }
\hskip2ex
\subfigure[$x_0=\sqrt{0.1}, t=1.7$]{
\includegraphics[scale=0.2]{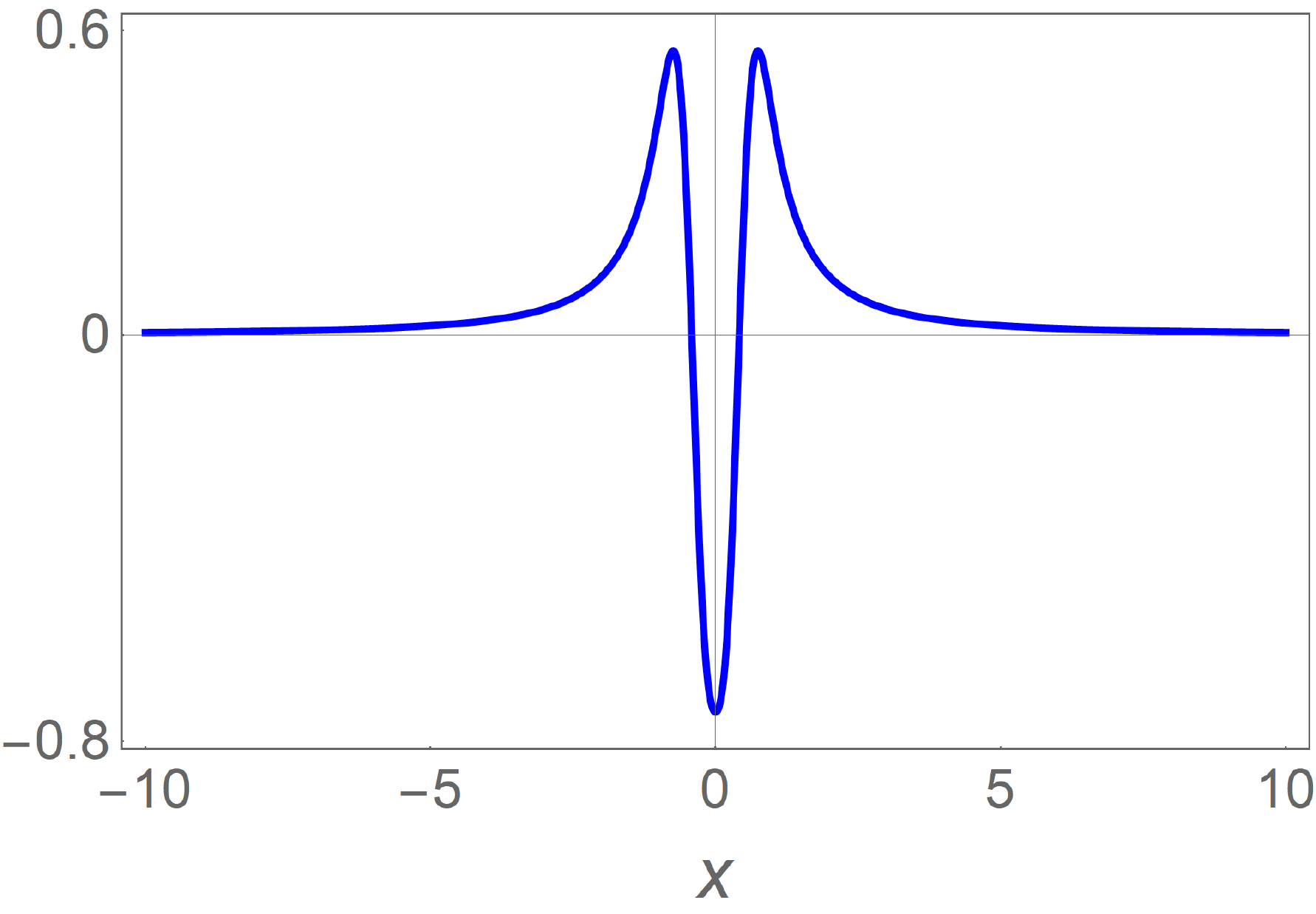} \label{VF-OR-10} }
\hskip2ex
\subfigure[$x_0=\sqrt{0.1}, t=5$]{
\includegraphics[scale=0.2]{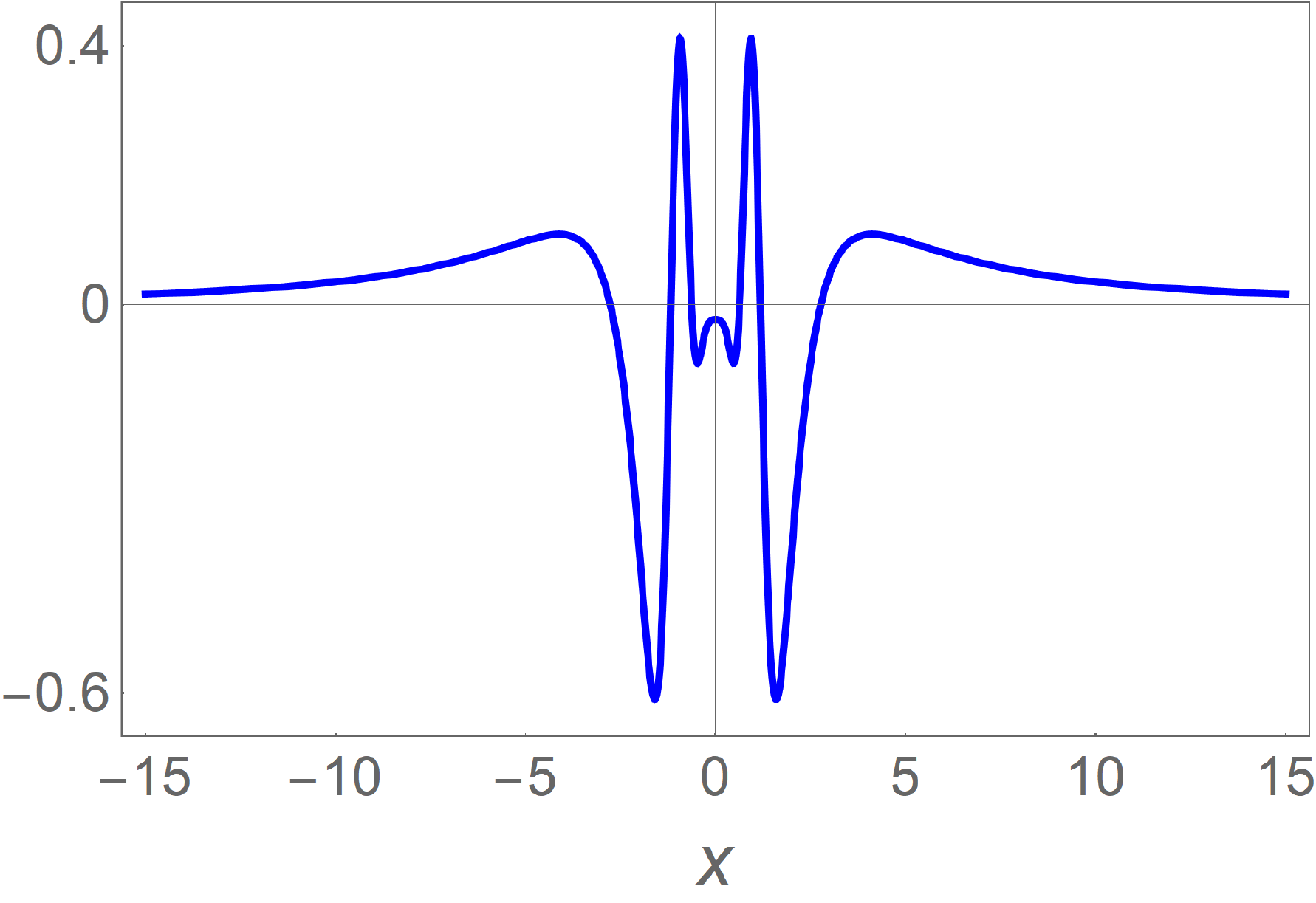} \label{VF-OR-11} }
\hskip2ex
\subfigure[$x_0=\sqrt{0.1}, t=6.5$]{
\includegraphics[scale=0.2]{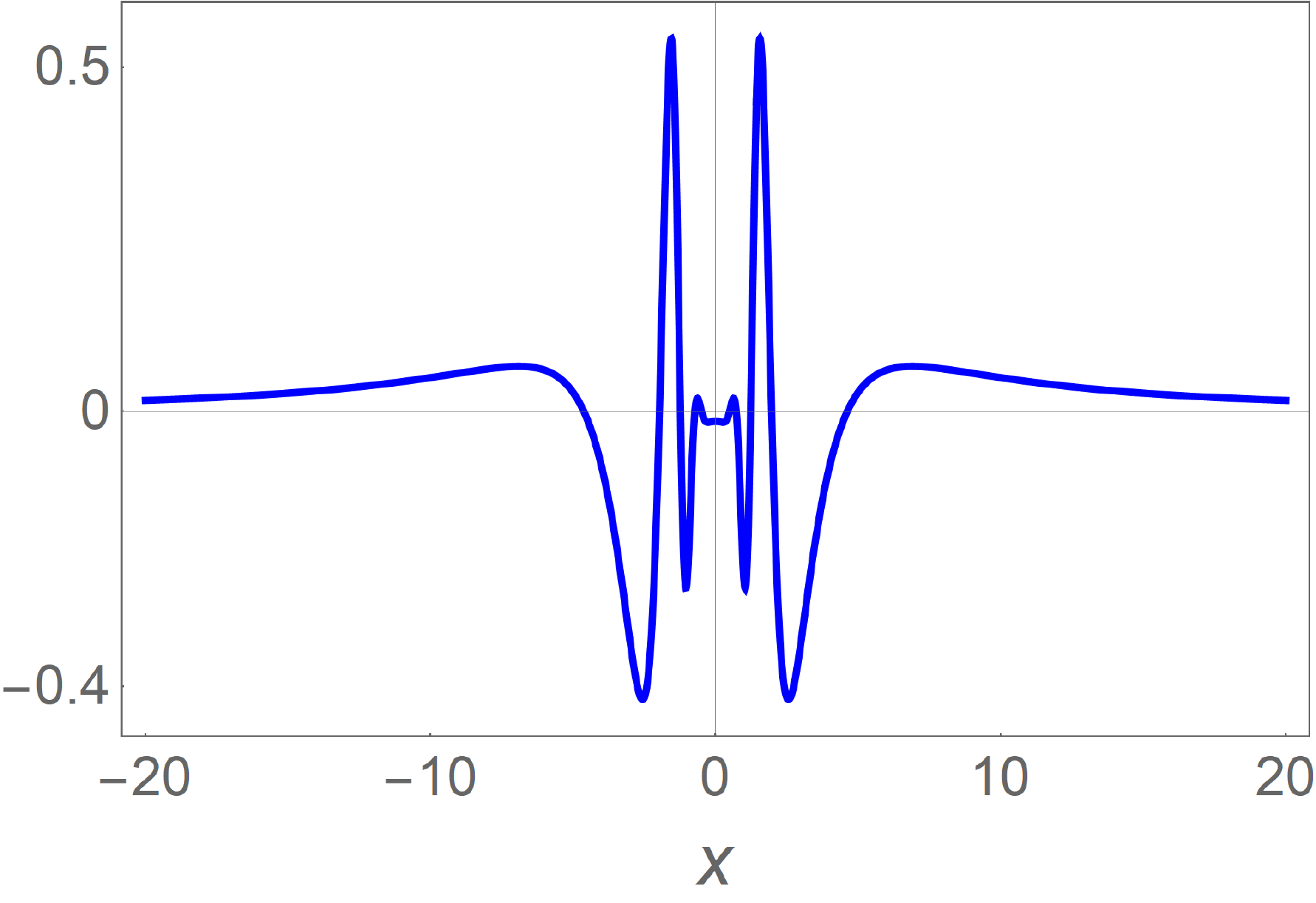} \label{VF-OR-12} }

\caption{Time-evolution of the function $u(x,t)$ defined in (\ref{propo})--(\ref{propo2}) with $\mu=\kappa=1$. The rows (characterized by the indicated values of $x_0$) show the emerging of zeros in $u(x,t)$ as $t$ increases. The columns exhibit the behavior of $u(x,t)$ as $x_0 \rightarrow 0$ (from top to bottom) at the indicated times.} 
\label{Fig1}
\end{figure}
\end{landscape}

The behavior of the solutions $u(x,t)$ defined in (\ref{propo})--(\ref{propo2}) is shown in the panel of Figure~\ref{Fig1} for $\mu = \kappa=1$. From top to bottom, the rows correspond to $x_0=1, \sqrt{0.5}, \sqrt{0.1}$. From left to right, the columns refer to $t=0.1,1.7,5,6.5$. An interesting profile of these functions is the emergence of zeros as time goes pass. The zeros arise in pairs at different times, they born superposed at $x=0$ and then propagate in opposite directions (symmetrically with respect to $x=0$). As the function $u(x,t)$ is initially a nonnegative pulse, the zeros are indeed nodes that propagate, together with the maxima and minima of the disturbance,  in wavelike form. For fixed values of $\mu$ and $\kappa$, the times at which we find new pair of nodes depend on the width of the initial Gaussian distribution. That is, they arise at shorter times for smaller values of $x_0$. We are interested in studying the behavior of such nodes as the disturbance $u(x,t)$ propagates. 

First, we use Theorems 1.2 and 1.4 of Ref.~\cite{Kil04} to rewrite $\theta_k$ as the absolutely convergent series \cite{Gor00}
\be 
\theta_k (x,t) =\frac{4^k}{\sqrt{\pi}}\left(\frac{\kappa t^2}{|x|}\right)^{1+2k}\sum_{\ell=0}^{\infty}\frac{(-1)^{\ell}\Gamma(\frac12+k+\frac{\ell}{2})}{\Gamma(1+4k+2\ell)\Gamma(-k-\frac{\ell}{2})}\left(\frac{\kappa t^2}{|x|}\right)^{\ell},
\label{solcom1}
\ee
where $x\neq 0$. The divergences of $\Gamma(-k-\frac{\ell}{2})$ eliminate the terms with even values of $\ell$ in the above expression, then
\be
u(x,t)  = \displaystyle\frac{\mu}{\sqrt{\pi}|x|} \sum_{k,n=0}^{\infty}\frac{(-1)^{k+1} }{k!}\left(\frac{x_0}{x}\right)^{2k} \left(\frac{2\kappa t^2}{|x|}\right)^{2n+1}  \lambda(n,k),
\label{u2}
\ee
with
\be 
\lambda(n,k)=\frac{\Gamma(1+n+k)}{\Gamma(3+4n + 4k )\Gamma \left( -\frac12- n- k \right)}. 
\ee
The latter formulae give us information about the nodes of the disturbance generated by the initial Gaussian-like perturbation defined in (\ref{gauss}). Of course, as the point $x=0$ has been omitted, the following description does not automatically hold for $\vert x \vert \leq  \epsilon$ as $\epsilon$ approaches to zero. 

The straightforward calculation shows that (\ref{u2}) can be rewritten in the form
\be 
u(x,t)= \frac{\mu}{\pi} \left( \frac{\kappa t^2}{x^2} \right) \sum_{s=0}^{\infty}\frac{\Gamma(2s+2)}{\Gamma(4s+3)}\left(\frac{\kappa t^2}{|x|}\right)^{2s} \Lambda_s(x_0,t), \quad x\neq 0,
\label{cor1} 
\ee
where $\Lambda_s(x_0,t)$ is the polynomial of $x_0t^{-2}$ given by
\be 
\Lambda_s (x_0, t)=\sum_{n+k=s}\frac{(-1)^n}{\Gamma(k+1)}\left(\frac{x_0}{2\kappa t^2}\right)^{2k}.\label{cor2} 
\ee

Now, let us analyze the series (\ref{cor1}) in terms of $\xi = \frac{\kappa t^2}{ \vert x \vert} >0$. For $\xi<<1$ we may consider the power with $s=0$ only. We have
\be
u(x,t) \approx \frac{\mu \kappa}{ 2 \pi} \left( \frac{t}{x} \right)^2, \quad  x \neq 0.
\label{aprox1}
\ee
The latter means that, no matter the value of $x_0$, the solution is free of zeros at short times. To illustrate the phenomenon, Figures~\ref{VF-OR-01}, \ref{VF-OR-05}, and \ref{VF-OR-09} show the behavior of $u(x,t)$ for the indicated values of $x_0$ at $t=0.1$.

At slightly larger times we may hold only the powers with $s=0$ and $s=1$. Thus, dropping the terms with $s\geq2$ we arrive at the expression
\be 
u(x,t) \approx \frac{\mu\kappa}{\pi} \left( \frac{t}{x} \right)^2 \left[\frac{1}{2}+\frac{\Gamma(4)}{\Gamma(7)}\left(\frac{\kappa t^2}{|x|}\right) \Lambda_1(x_0,t)\right], \quad x\neq 0, 
\label{ut2}
\ee
where
\be 
\Lambda_1(x_0, t)=-1+\left(\frac{x_0}{2\kappa t^2}\right)^2. 
\ee
Given $x_0 \geq 0$, the values of $t$ such that $\Lambda_1<0$ permit the presence of a pair of zeros  in function (\ref{ut2}). Before such values, the function $u(x,t)$ exhibits a global minimum that is positive and goes to zero as $t$ increases. Then the minimum becomes equal to zero (the time at which the first pair of nodes is created, both superposed at $x=0$), and finally it takes negative values (the nodes start to propagate in opposite directions with respect to $x=0$). The second column (from left to right) of Figure~\ref{Fig1} shows the situation in which the minimum of $u(x,t)$ is negative for three different values of $x_0$. Although the three graphics are evaluated at $t=1.7$, notice that the minimum is as deep as $x_0$ is short. The latter shows that the positions of the nodes at a given time depend on $x_0$.

At larger times, the value of $x_0$ determines the number of zeros as well as their distribution. For example, in the third and fourth columns (from left to right) of Figure~\ref{Fig1} we appreciate that the number of nodes increases as $x_0$ decreases at a given time. In general, such number increases as $\xi \rightarrow  \infty$. Then, the time $t$ at which a new pair of nodes arises is shorter for smaller values of $x_0$. Remarkably, at the limit $x_0 \rightarrow 0$, for the polynomial (\ref{cor2}) we have 
\be 
\lim_{x_0 \to 0} \Lambda_s(x_0, t)=(-1)^s.
\label{cor3} 
\ee
From (\ref{cor1}) and (\ref{cor3}) one has
\be 
\lim_{x_0 \rightarrow 0} u (x,t)= \frac{\mu}{\pi} \left( \frac{\kappa t^2}{x^2} \right) \sum_{s=0}^{\infty}(-1)^{s} \frac{\Gamma(2s+2)}{\Gamma(4s+3)}\left(\frac{\kappa t^2}{|x|}\right)^{2s}, \quad x\neq 0,
\ee
which corresponds to the series expansion of $u^{(\delta)} (x,t)$ reported in \cite{Oli18}.

\section{Concluding remarks} \label{sec:4}

We have shown that the (modified) complementary equation $u_{tt} -\kappa \mathbb D u=0$ can be solved in analytic form by considering the Cauchy problem for zero initial velocity and the Gaussian distribution as initial disturbance. The solutions exhibit nodes that arise in pairs at different times and propagate from $x=0$ in wavelike form. The number of zeros in a given time-interval increases as the width of the distribution is reduced. At the very limit in which the width becomes equal to zero we recover the solutions to the Cauchy problem with the initial disturbance as a Dirac delta pulse. The possible physical applications of the modified complementary equation represent an open problem, which we shall face elsewhere.

\section*{Acknowledgment}

Financial support form Ministerio de Econom\'ia y Competitividad (Spain) grant number MTM2014-57129-C2-1-P, Consejer\'ia de Educaci\'on, Junta de Castilla y Le\'on (Spain) grant number VA057U16, and Consejo Nacional de Ciencia y Tecnolog\'ia (Mexico) project number A1-S-24569, is acknowledged.


\end{document}